# Optimizing recording speed and interrogation window for rotating flow recorded in the ambient light: PIV analysis

Shailee P Shah[1], Nayan Mumana[1], Preksha Barad[1], Rucha P Desai[1#], Pankaj S Joshi[2]

[1] Department of Physical Science, P D Patel Institute of Applied Sciences, Charotar University of Science and Technology, CHARUSAT Campus, Changa – 388 421, Gujarat, India.

[2] International Centre for Cosmology, Charotar University of Science and Technology, CHARUSAT Campus, Changa – 388 421, Gujarat, India.

[#] E-mail address for correspondence: ruchadesai.neno@charusat.ac.in

**Abstract**

The present study reports PIV analysis of the surface flow profile using a smartphone camera in ambient light instead of high-tech equipment like a professional camera and high-power laser/ LEDs. Additionally, it provides a stepwise method for optimizing recording speed and interrogation window size for the vortex flow generated at different rotational frequencies of the magnetic stirrer. The optimization method has been explained with an example of the vortex flow generated by a magnetic stirrer. The analysis has been carried out using the Matlab-based application PIVlab. Finally, the optimized parameters have been compared with the Burger vortex model, which shows good agreement with the PIV data. The proposed method can also determine the sureface flow of opaque liquids.

## 1.Introduction

Flow measurement and visualization are inseparable parts of fluid mechanics. Visualization of simple laminar flow to complex turbulent flow is possible due to analytical and computational fluid flow simulations. However, experimental measurements and visualization are equally crucial for the application and educational point of view. Particle Image Velocimetry (PIV) is well- established and vastly used technique for the experimental velocity estimation of fluid flow (Raffel et al., 2007). A typical PIV setup consists of a high-power multi-pulsed leaser sheet (to illuminate tracer seeded flow), a high-speed camera (to record the motion of the illuminated flow), a synchronizer (to synchronize laser pulse with a camera), and additional optical components & its arrangements (to convert laser beam into a sheet). In this method, initially, the flow geometry can be traced by the tracer particles illuminated by a laser sheet, and subsequently, flow geometry can be captured by the high-speed camera. Later, the captured images/ video can be processed by the standard PIV software. Finally, the software estimates the velocity based on the displacement of tracers using cross-correlation between two frames (Raffel et al. 2007)(2016). PIV, a non-destructive imaging-based technique, has gained interest among the scientific community, researchers, and academicians. Moreover, the technique can be applied to other fields, such as oceanography, marine biology, zoology, and microbiology(Raffel et al., 2007)(Minichiello et al., 2020).

Despite having advantages, PIV techniques have several limitations say the need for trained human resources, proper maintenance, space, and safety hazards associated with a high-power pulsed laser (class 4, >500mW), and above all, the high cost of all the instruments involved (Minichiello et al. 2020). All these factors force educational institutions to restrict their use for new learners. On the other hand, various commercial PIV systems have been developed for educational purposes (e.g., THERMOFLOW, ePIV, HEMOFLOW, MiniPIV) (Minichiello et al. 2020) to enable users the ease of operation with limited variable components. For example, users can only change the inlet and outlet parameters, water level, and seeding density. Additionally, scientists and technologists have been working towards developing a PIV system with the motto of ease of operation, using advanced technology with slick models of equipment with less space utilization, economical, and minimal maintenance.

In PIV analysis, significant technological advancement has provided alternatives for the high-tech equipment, e.g., high-power pulsed lasers can be replaced with high-power LEDs and high-speed professional cameras with smartphone cameras, which in turn makes the overall PIV system safe, simple, easy to operate, and notably economical (Kashyap et al.; Buchmann et al. 2010; Willert et al. 2010; Harshani et al. 2015; Hain et al. 2016; Dai et al. 2017; Aguirre-Pablo et al. 2017). Tomographic PIV, 3D PIV, mIPIV, and smartPIV are a few examples of technological developments. Tomographic PIV uses multiple smartphone cameras and different colored LEDs (Aguirre-Pablo et al. 2017), while the 3D PIV technique uses a single camera and structured light illumination (Aguirre-Pablo et al. 2019). Both these studies show good agreement with the standard stereo-PIV data for the same flow. Simultaneously, mobile-based applications for PIV measurement, such as mIPIV and smartPIV, have been developed from the educational perspective (Minichiello et al. 2020; Cierpka et al. 2021). It provides the live visualization of flow, allows users to capture the flow at various fps, and gives the data as a text file and vector image (Cierpka et al. 2021). In all these papers, either high-power lasers or high-power LEDs have been used. Contrary, the present work focuses on using ambient light for the water vortex generated using a magnetic stirrer.

Additionally, the effect of recording speed with different frames per second (fps), i.e., 30, 120, and 240 fps, has been illustrated here using the smartphone camera. Most publications mainly report either a trial-and-error method for selecting an interrogation window (IW) or the results without specifying the effect of IW. However, IW is an integral part of the process which directly affects the final output; it should be adequately addressed. Hence, here we have given a stepwise process for selecting the interrogation window (IW). The optimization step is critical learning for beginners. Here, we followed ITTC guidelines (2016) to optimize the interrogation window size. Results have been analyzed using the MatLab-based open-source extension PIVlab (Thielicke and Stamhuis 2014).

## 2. Experimental

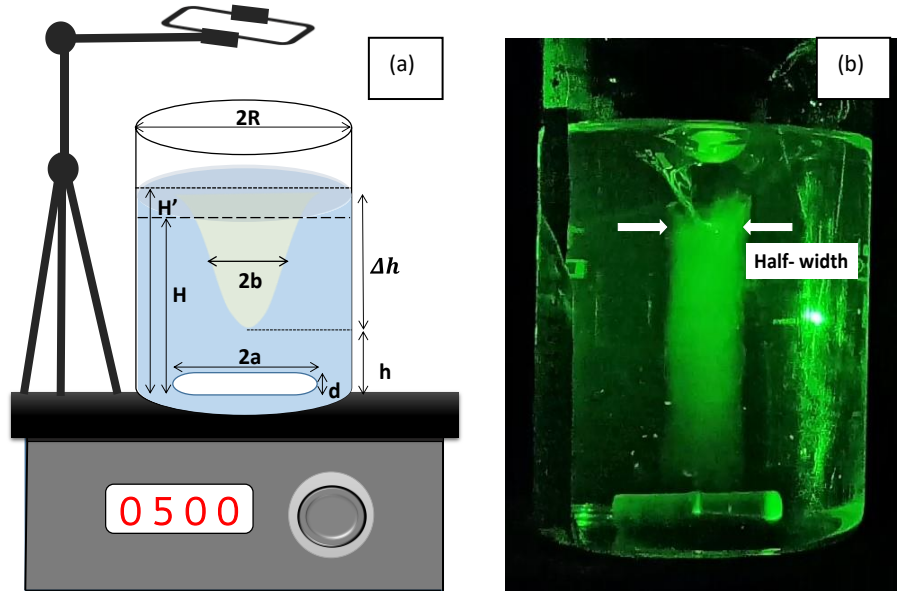

Figure 1 (a) Schematic representation of experimental setup, (b) Vortex generated using magnetic stirrer. A half width of the vortex has been determined based on the cylindrical wall of dye.

A water vortex was generated in a cylindrical glass beaker using a magnetic stirring bar. Figure 1(a) shows a schematic of the experimental setup for generating the water vortex. The dimensions of the vessel and the stirrer bar are given in Table 1. The experiments have been carried out for different rotational speeds (ω) i.e., $8.33 s^{-1}$, $10.00 s^{-1}$, $11.66 s^{-1}$, and $13.33 s^{-1}$ of the magnetic stirrer bar. Here, the vortex depth was measured experimentally using a ruler.

Simultaneously, the surface flow has been captured from the top of the vessel using a mobile camera (refer to Table 2) with different recording speeds, i.e., 30, 120, and 240 fps (see Figure 1(a)). The position of the mobile phone was fixed throughout the experiments.

Table 1 Dimensions of a vessel and magnetic stirrer bar

| **Dimensions** | |
|---|---|
| The diameter of a vessel (2R) | 0.105 ± 0.001 m |
| Height of vessel | 0.150 ± 0.001 m |
| Filled water at a height (H) | 0.110 ± 0.001 m |
| Length of the stirrer bar (a) | 0.020 ± 0.001 m |
| Width of the stirrer bar (d) | 0.009 ± 0.001 m |

Next, the half-width of the vortex was measured by injecting the fluorescent dye near the center to see the illuminated cylindrical vortex wall (Figure 1(b)). The radius of the cylindrical wall is equivalent to the critical radius ($r_c$) or a half-width (2b) of the vortex (Halász et al. 2007). To measure the $r_c$ an image of the vortex was captured and later analyzed using PIVlab. The scale was initially calibrated during the PIV analysis using the calibration stick by entering the known distance (in meters). Once the calibration is done, PIV uses this scale for further analysis. In the present experiment, the vortex half-width ($r_c$ or 2b) determined from PIV was 0.0094±0.0002 m. The apparent advantage of PIV analysis is its accuracy compared to visual observation.

## 2.1 Data extraction from the PIVlab

The videos were recorded at different fps using the "One Plus Nord 2" mobile phone camera. Camera features are given in Table 2.

Table 2 Camera features

| Model Name | OnePlus Nord-2 |
|---|---|
| Camera Sensor | Sony Exmor IMX766 |
| Pixel Size | 1.0 µm |
| Lens Quantity | 6p |
| MP | 50 MP |
| OIS, Optical Zoom, Autofocus | YES |
| Aperture | f/1.88 |
| Video recording | • 4K at 30 fps<br>• 1080p at 30/60 fps<br>• 720p at 30/60 fps |
| Super slow-motion video recording | • 1080p at 120 fps<br>• 720p at 240 fps |

Figure 2 shows the flowchart to perform a PIV analysis of the recorded flow followed by its description. In the present work, (i) 50 consequently frames were inserted from the video, and (ii) the three pass interrogation windows with 50% overlapping with (a) 128 x 64 x 32pixels and (b) 64 x 32 x 16 pixels were used.

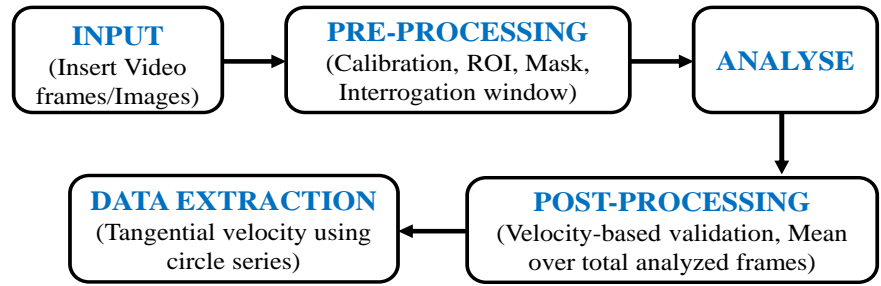

Figure 2 flowchart for data extraction from PIVlab

**Step 1: Input**

Inset video frames with the desired number of frames or inset images (minimum two) to detect the displacement of tracer particles.

**Step 2: Pre-processing**

Pre-processing allows the selection of the region of interest (ROI), masking of unnecessary areas/ objects in the images, and image calibration into the physical unit. The path for the same is as follows.

Go to the menu bar and select,

- Image settings > Exclusions (ROI, mask) > select ROI > draw mask > apply current mask to all frames
- Calibration > select reference distance in image > enter real distance (mm) & time steps (ms)

Here the time step is 1/fps.

**Step 3: Analyze**

The interrogation window (IW) size should be defined before applying the analyze command. The process enables software to generate velocity vectors from the displacement of the tracer particles.

Go to the menu bar and select,

- Analysis > PIV settings > enter interrogation window size (unable necessary pass with the desired overlapping according to flow) > Analyze all frames.

**Step 4: Post-processing**

Once the analysis process is completed, one can remove the random vectors automatically by introducing velocity limits. Alternatively, manually individual vectors can be removed. The steps for the same are as follows.

Go to the menu bar and select,

- Post-processing > velocity-based validation > select velocity limits > apply to all frames

The mean flow can be determined based on the total inserted frames as given below.

- Plot > Temporal: Derive parameters > calculate mean

**Step 5: Data extraction**

The software gives the data of u and v components of velocity vectors, tangential velocity, vorticity, and shear rate. In the present work, our interest is to extract the tangential velocity component only. For this,

Go to the menu bar and select,

- Extractions > parameters from polyline > Type > circle series > plot data > save result as a text file

## 3. Result and discussion

The study has been conducted in two parts to understand the significance of recording speed and interrogation window, i.e., (i) varying fps viz. 30fps, 120fps, and 240fps while keeping the interrogation window constant, (ii) varying interrogation window (IW) size for a single video recorded at a given fps.

### 3.1 Optimization of recording speed (fps- frames per second)

**Error! Reference source not found.** a(i, ii, iii) illustrates typical tangential velocity ($v_\theta$) distribution obtained from the PIV for respectively 30fps, 120fps, and 240fps at a constant rotations frequency of magnetic stirrer ($\omega$) 11.66s$^{-1}$. Here IW area selected was 128 x 64 x 32 pixels. It is evident from the figure that the number of vectors observed is less at 30 fps, maximum at 120 fps, and moderate at 240 fps for a constant $\omega$= 11.66 s$^{-1}$. It also noted that the velocity magnitude increases with increasing fps. Detail vortex flow geometry can be determined precisely at 240 fps. The presently used mobile camera has a maximum of 240 fps, while one could record video at higher fps. Figure S1 shows the effect of varying fps, i.e., 30, 60, 120, 240, and 960 fps for $\omega$ = 3.33 s$^{-1}$. It is inferred from the figure S1 that, the $v_\theta$ increases with increasing fps reaching maxima at 240 fps, and then further increase in fps (i.e., 960 fps) results into decrement in the $v_\theta$. The fact that increasing fps reduces the resolution, which may hamper the pixel identification, result in generating a false vector in PIV.

**Error! Reference source not found.** b(i-iii) shows a typical tangential velocity ($v_\theta$) data obtained as a function of radial distance (r) from the central axis of the vortex at frequencies ($\omega$) 8.33s$^{-1}$, 10.00s$^{-1}$, 11.66s$^{-1}$, and 13.33s$^{-1}$. As shown in Figure b(i), the magnitude of $v_\theta$ is negligible at 30 fps for different $\omega$. The velocity distribution (see Figure 3a(i)) and the tangential velocity plot (see Figure 3b(i)) at 30fps show negligible velocity magnitude. It means the tracers leave the IW in the next frame, resulting in the false correlation peak. Therefore, 30fps recording speed is insufficient for the rotational frequencies 8.33s$^{-1}$, 10.00s$^{-1}$, 11.66s$^{-1}$, and 13.33s$^{-1}$. The nature of tangential velocity obtained at 120 and 240 fps is similar; however, a higher magnitude is observed at 240 fps. It is known that the velocity magnitude increases with increasing $\omega$. When the flow was recorded at 120fps, for ($\omega$) 8.33s$^{-1}$, 10.00s$^{-1}$, 11.66s$^{-1}$, and 13.33s$^{-1}$, velocity magnitude increased, but for $\omega$ 13.33s$^{-1}$, it decreased. The data recorded at 120 and 240 fps have been normalized by $\omega$ to understand the effect of recording speed (fps).

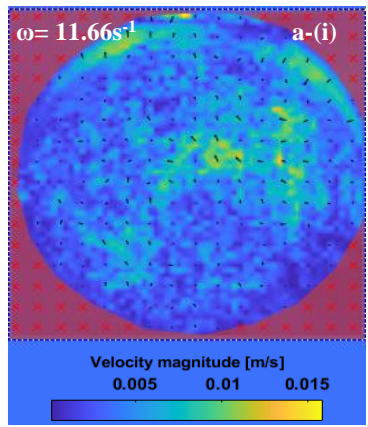 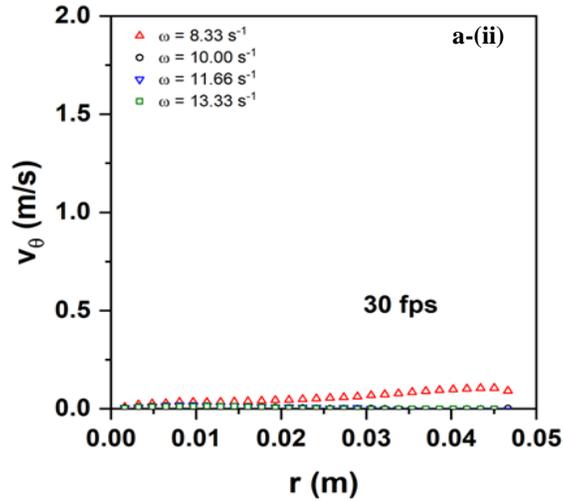
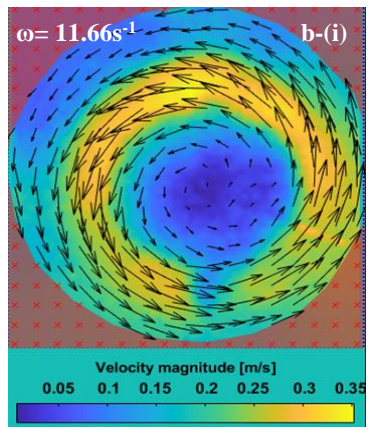 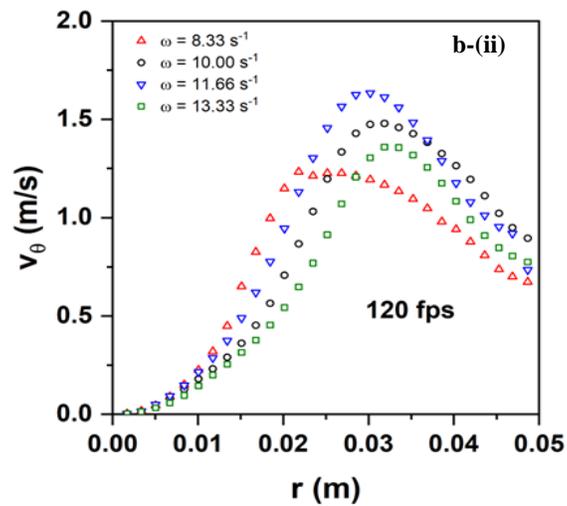
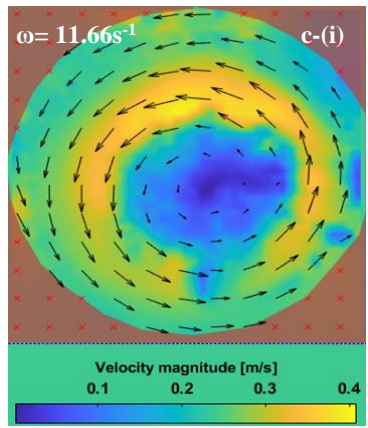 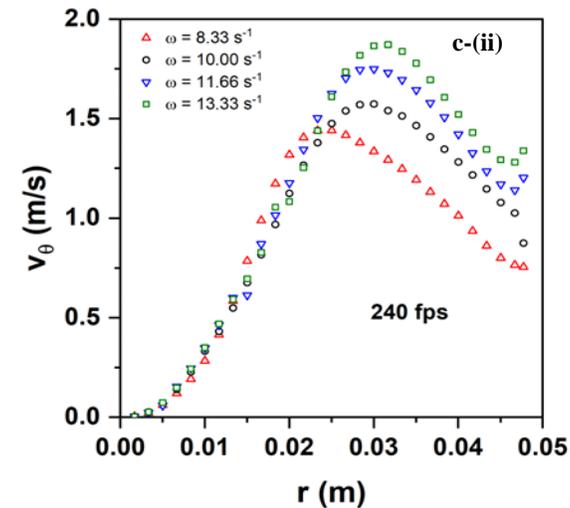

Figure 3(a&b): a (i – iii) shows typical tangential velocity ($v_\theta$) distribution for rotations frequency of magnetic stirrer ω= 11.66s-1 @ 30fps, 120fps, and 240fps respectively (IW: 128 x 64 x 32pixels). b (i - iii): Variation of the tangential velocity ($v_\theta$) as a function of beaker radius (r) for different ω at (i) 30fps, (ii) 120fps, (iii) 240fps

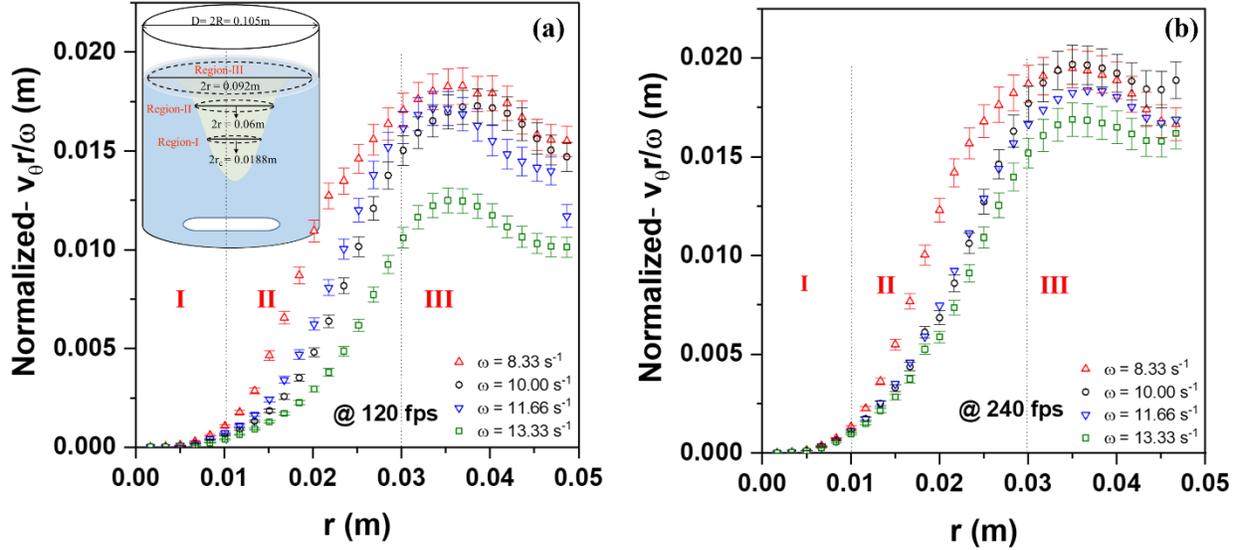

Figure 4 Normalized tangential velocity for (a) 120 fps and (b) 240 fps. Inset: Graphical illustration of regions of vortex

Figure 4 (a & b) shows a normalized tangential velocity plot at 120 and 240 fps respectively for IW: 128 x 64 x 32 pixels. Inset of Figure 4 is a graphical illustration of three vortex regions, which are explained below. Both Figures 4 (a & b) are virtually divided into three regions, i.e., (I) from 0 m to 0.01 m, (II) 0.01 m to 0.03 m, and (III) 0.03 m to 0.047 m, which is the radius of the beaker.

It is evident that all data superimpose on each other in the region (I). The region-I is also referred the critical radius ($r_c$) (also known as vortex core or half-width(2b))(see Figures 1(a & b) & inset of Figure 4(a)). The $r_c$ is the radius up to which the vortex obeys rigid rotation irrespective of the stirring speed. As mentioned above $r_c$ =0.0094±0.0002 m, the superimposition of the region (I) is attributed to the rigid rotations. In particular, the half-width can be expressed by $b = \frac{\beta a v}{d(gR)^{1/2}}$, where β is the proportionality constant, ν is the kinematic viscosity of the liquid, and g is the gravitational constant (Halász et al. 2007). This equation depicts that the half-width depends only on the vessel's radius and the stirrer's dimensions, while it does not depend on the rotation rate and the water level. Thus, half-width(2b) would be the same for each rotational speed of the stirrer bar.

Region II experiences the maximum effect of the stirring bar, including down-welling and up-welling. Therefore, it leads to observe deviation due to variation in ω. At 120 fps, a relatively large deviation is observed for all the ω. While at 240 fps, most of the data overlap (within the error) for all the ω except ω = 8.33 s$^{-1}$. Region III is far away from the vortex core and wall of the beaker, where one can expect superimposition of data due to the minimum effect of the stirring bar and wall. In this region, data recorded at 240 fps (see Figure 4(b)) superimposes while deviation is observed at 120 fps (Figure 4(a)). By comparing the data of Figure 4 (a & b) it is concluded that data recorded at 240 fps are more promising than at 120 fps. Hence, the experimental data of 240 fps is only considered to understand the effect of IW.

### 3.2 Optimization of interrogation window size

The software divides inserted frames into the small sub-images called interrogation window (IW), which performs correction correlation for each window to generate velocity vectors. Tracer particles belonging to the same interrogation window are assumed to have the same local velocity. The selection of interrogation window size usually depends on the size and number density of the tracer particles. As per the ITTC guidelines, there should be a minimum of 10 tracer particles in a single interrogation window to avoid the random displacement correlation peak (2016).

When the IW size is too larger, tracer displacement in the same window can result in the false magnitude of the correlation peak.

In contrast, small IW compares to the tracer particle size results in the clipping of tracers at the edge of the window. Because of the clipping, the random error increases as the average number of particles in a single window reduces (2016). Therefore, the optimization of IW size is necessary. Instead of one IW, three pass IW was used in the present analysis with 50% overlapping. Here, a single video was analyzed to optimize IW size for $\omega = 8.33s^{-1}$ and $13.33s^{-1}$ at 240fps. Table 3 shows the two combinations of the IW, in which one has a primary IW of 64 pixels, and the other has 128pixels.

Table 3 combination of interrogation window size

| Pass | The interrogation window size in pixels | |
| --- | --- | --- |
| 1 (Primary IW) | 64 | 128 |
| 2 (Middle IW) | 32 | 64 |
| 3 (Final IW) | 16 | 32 |

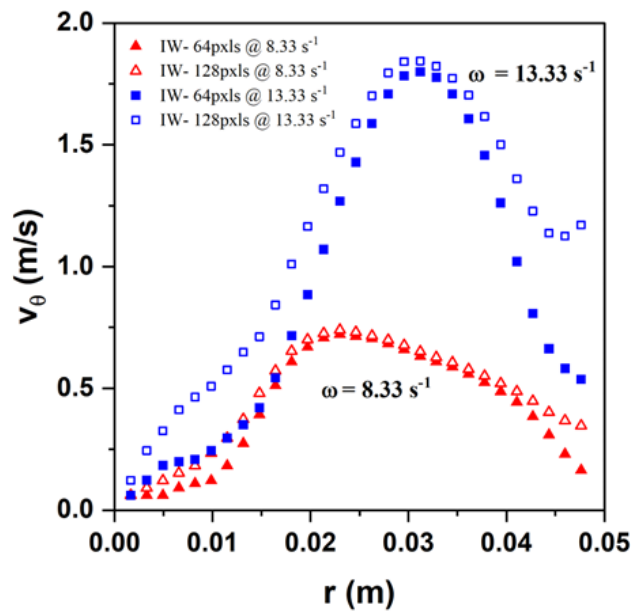

Figure 5 Tangential velocity plot for IW 64pxls and 128pxls

Figure 55 depicts the data of tangential velocity for $\omega = 8.33s^{-1}$ and $13.33s^{-1}$ at 240fps. The overlapping of the data for $\omega = 8.33s^{-1}$ indicates the reliability of the combination of IW. Whereas, for $\omega = 13.33s^{-1}$, the large window sizes (128pixels) seem more accurate. When the rotational speed of the stirrer increases, the jet flow near the center of the vortex becomes stronger, resulting in the inhomogeneous distribution of the tracer particles. On the contrary, due to jet flow, fewer particles will be observed near the wall of the vessel. In Figure 5, tangential velocity increases with increasing radius reach to maxima followed by decrement. The fact that vortex mouth (diameter) increases with increasing stirring speed. However, after reaching maximum r, the decrement in the tangential velocity is attributed to the wall effect. However, a large decrement in the velocity magnitude at 64 pixels could be due to the coverage of fewer tracer particles, leading to false velocity magnitude. In comparison, 128 pixels cover more particles which represent more realistic data. Hence, it is concluded that 128 x 64 x 32pixels IW with 50% overlapping is optimum for the entire range of $\omega$.

## 3.3 Confirmation of velocity magnitude

As mentioned above (inset Figure 4(a)), a vortex region that is far away from the vortex core and beaker wall generally follows the Burger vortex model (Halász et al. 2007). The same model can also fit whole data when the limiting factors like wall effect and perturbation created by the stirrer do not affect the vortex. However, as mentioned above, in the present work, considering the beaker dimensions and the stirring speed, regions I & II are excluded from the fit. The tangential velocity component of the Burger vortex can be expressed by equation 2 (Halász et al. 2007).

$$v_\theta = \frac{C}{r}(1 - e^{-r^2/r_c^2}) \qquad (2)$$

Where C is the vortex strength. It can be determined from $C^2 = \Delta h g r_c^2 / ln2$, where g is the gravitational constant (= 9.8 m/s), Δh is the change in the height of the vortex, $r_c$ is the critical radius of the vortex. The vortex strength calculated experimentally is denoted here as $C_{cal}$. Similarly, the tangential velocity ($v_\theta$) data compared to equation (2) with the variables C and $r_c$ are denoted here as $C_{fit}$ and $r_{c(fit)}$, respectively.

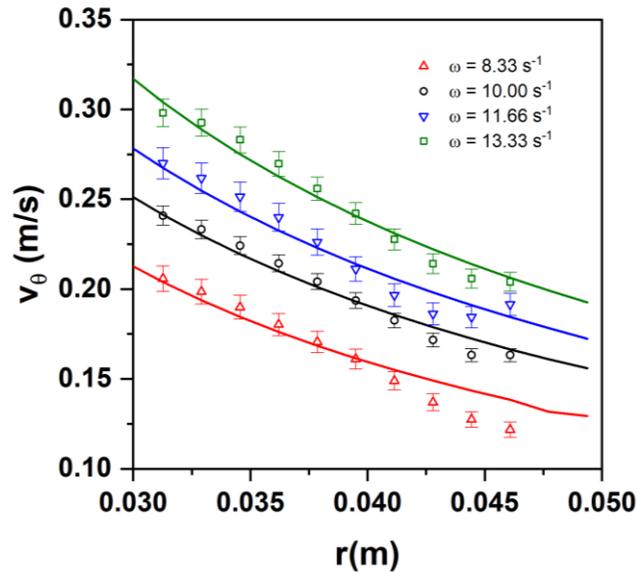

Figure 6 The tangential velocity ($v_\theta$) data extracted from PIV (open symbol) fitted to the Burger vortex model (solid line) at various rotational frequencies (ω)

Figure 6 shows the tangential velocity ($v_\theta$) data extracted from PIV (open symbol) fitted to the Burger vortex model (equation 2) (solid line) as a function of radial distance (region – III) at various rotational frequencies (ω). It is evident from Figure 6 that the overserved $v_\theta$ agrees with the Burger vortex model. The parameters obtained from the fit are shown in Table 4. As shown in Table 4, the $r_{c(fit)}$ = 0.0095 ± 0.0002 m matches that of experimentally observed $r_c$ (0.0094 ± 0.0002 m). Table 4 also confirms the agreement of $C_{cal}$ and $C_{fit}$ for the different values of Δh. It also validates the proposed method of optimizing the parameters. The present method can also be applied to study the surface flow of opaque liquids.

Table 4 calculated ($C_{cal}$) and fitted ($C_{fit}$) vortex strength at different stirring speeds (ω)

| Stirring speed ω ($s^{-1}$) | Δh (m) | $C_{cal}$ ($m^2$/s) | $C_{fit}$ ($m^2$/s) | $r_{c(fit)}$ (m) |
|---|---|---|---|---|
| 8.33 | 0.024 ± 0.001 | 0.0055 ± 0.0015 | 0.0063 ± 0.0002 | 0.0095 ± 0.0002 m |
| 10.00 | 0.042 ± 0.001 | 0.0073 ±0.0018 | 0.0079 ± 0.0002 | |
| 11.66 | 0.053 ± 0.001 | 0.0081 ± 0.0020 | 0.0089 ± 0.0003 | |
| 13.33 | 0.072 ± 0.001 | 0.0096 ± 0.0022 | 0.0095 ± 0.0002 | |

## 4. Conclusion

Optimization of recording speed (fps) and interrogation window size for rotating flow recorded in the ambient light using PIV has been carried out here. In the experimental condition, the optimized parameters are 240 fps and 128 x 64 x 32 pixels for the video recording speed and interrogation window size, respectively. The tangential velocity determined based on the optimized condition matches that of the Burger vortex model, which validates the presently described method. As the present method does not require high-tech equipment like high-power lasers/LEDs, or professional cameras, budding researchers, students, and industries can use PIV easily for vortex analysis. Furthermore, the method can also be used to analyze the surface flow of opaque liquids.

## Acknowledgment

SPS acknowledges the education department, Gujarat state, India, for providing SHODH fellowship vide no. 202001300006. Part of the work has been carried out under postgraduation dissertation project by NM and PB.

# Supplementary Information

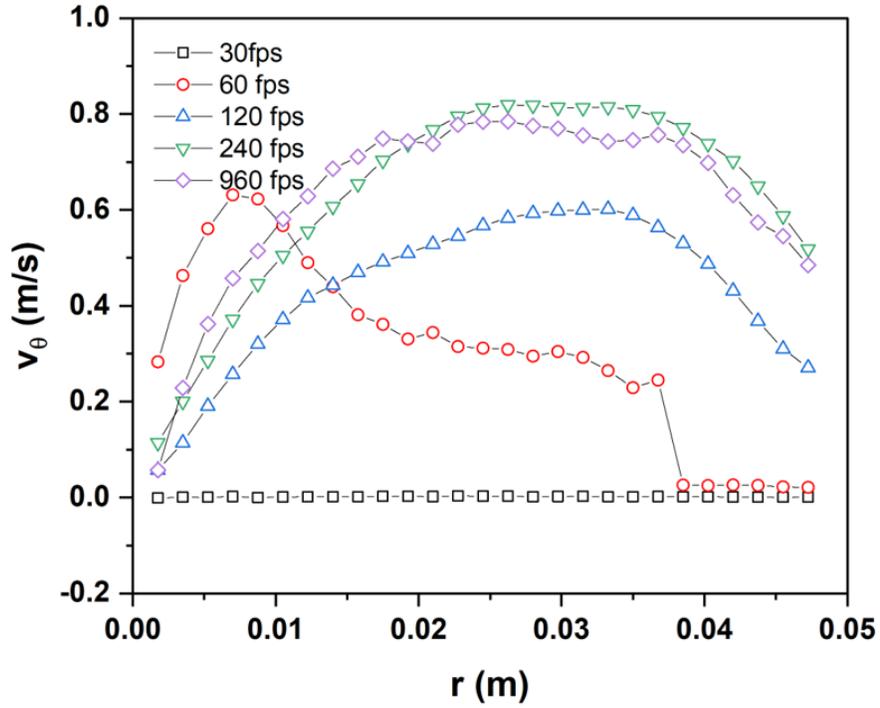

**Figure S1**: Variation of tangential velocity ($v_\theta$) as a function of radial distance (r) for various recording speed i.e., 30, 60, 120, 240, and 960 fps for $\omega = 3.33$ s$^{-1}$